\documentclass{aa}
\usepackage{graphicx}
\def\approxgt{\mathrel{\hbox{\rlap{\lower.55ex \hbox {$\sim$}}
        \kern-.3em \raise.4ex \hbox{$>$}}}}
\def\approxlt{\mathrel{\hbox{\rlap{\lower.55ex \hbox {$\sim$}}
        \kern-.3em \raise.4ex \hbox{$<$}}}}

\begin{document}
   \title{Dimensional Analysis applied to Spectrum Handling in Virtual Observatory Context}

   \author{Pedro Osuna, Jesus Salgado}

   \offprints{P. Osuna, J. Salgado}

   \institute{ESAVO, European Space Astronomy Centre, ESA\\
              Apartado 50727, E-28080 Madrid, Spain \\
              \email{Pedro.Osuna@esa.int,Jesus.Salgado@esa.int}
              }

   \date{}

   \abstract{
The handling of units in an automated way by software systems can
be a cumbersome procedure when the units are parsed as strings.
Software systems parsing units have to take into account extensive
tables of unit names, not always identical between different
standards. In the Virtual Observatory context, the transfer of
metadata in access protocols is specially critical for the
understanding of the content of the data (many of which are legacy
data). Driven by this issue, we present a way to handle units
automatically which is based in dimensional analysis
considerations. Although the approach presented here can be valid
to any other branch of physics, we concentrate mainly in the
application to spectrum handling in the VO context, and show how
the proposed solution has been implemented in the VOSpec, a tool
to handle VO-compatible spectra.
    \keywords{
   Dimensional Analysis, Spectrum, Virtual Observatory, Unit, WCS
            }
            }

\authorrunning{Osuna-Salgado}

\titlerunning{Dimensional Analysis and VO Spectra}

   \maketitle
%

\section{Introduction}
Handling of units in spectra can be cumbersome when dealing with
different Flux energy densities. For instance, the conversion
between $F_{\lambda}$ and $F_{\nu}$ is normally done by parsing
the unit name strings. Despite several efforts to homogenize unit
name strings (See \cite{GREISEN2004},  \cite{TAYLOR1995},
\cite{GEORGE1995}), several different standards do exist, forcing
the parsing-string mechanism to stick to one or several of those
standards. On top of this, the adaptation of already existing
legacy data to one or other standard might be a cumbersome -and
even sometimes impossible- task.

The aforementioned problems led us to try and figure out a way to
automatically handle units without the necessity to parse strings.
Using dimensional analysis, we have devised an algorithmic way to
convert between dimensionally homogeneous quantities. As a simple
example, the string-parsing algorithm needed to convert between
$W/cm^{2}/\mu m$ and $erg/cm^{2}/s/\AA$ (both $F_{\lambda}$) would
be substituted for an algorithm to go from $10^{10} M L^{-1}
T^{-3}$ to $10^{7} M L^{-1} T^{-3}$, i.e., dividing by a factor of
$10^{3}$.

In section 2 we give a general mathematical formalism of the
relevant parts of Dimensional Analysis techniques that will be
needed for the handling of this problem. In this section we will
follow with very slight modifications the excellent book by
\cite{SZIRTES1997}. We also describe the handling of the
dimensional matrix to unveil how to extract dimensional relations
between different units on the same problem. In section 3 we apply
the theory to the case of $F_{\lambda}$ and $F_{\nu}$ conversions.
And in section 4, we give a general algorithmic method for the
conversion between different unit systems for the case of spectra
together with details about the use of this technique in the
VOSpec, a tool to handle SSAP (Simple Spectrum Access Protocol)
compatible spectra developed at the European Space Astronomy
Centre (ESAC) of the European Space Agency.

\section{Dimensional Analysis overview}
Dimensional analysis helps in the understanding of certain
problems for which no analytic mathematical formulation exists, or
for which the mathematical formulation is too complex. In these
cases, the dimensional analysis allows to extract certain
conclusions about the behavior of the system without the need of
specific mathematical formulae relating the different variables of
the problem at hand.

In physics we deal with quantities which have certain dimensions.
These are combinations of a reduced number of basic or fundamental
dimensions. These fundamental dimensions form a dimensional
system. Dimensional systems can range from mono-dimensional (only
a fundamental dimension is used to represent any physical
quantity) to omni-dimensional (all dimensions are fundamental
dimensions)(See ref \cite{SZIRTES1997} for examples). The
intermediate, and mostly used, is the multidimensional system, in
which a reduced set of fundamental dimensions is used. In the
examples that will follow, we will adhere to the most widely used
system of MASS-LENGTH-TIME system with the addition of
Temperature, Electric current and luminosity to deal with other
more complex problems. Following a long tradition in dimensional
analysis (see \cite{MAXWELL1890}) will call M-L-T these
fundamental dimensions. Following this convention, a physical
quantity, e.g., a Force, would be represented as follows:

$[F] = [m] [a] = M L T^{-2 }$

where square brackets should be read as "dimensions of" and we
will call the rightmost part the "dimensional equation" of the
quantity under consideration.

\subsection{Dimensional Matrix}
Let $F(V_{1}, V_{2}, ..., V_{n})=0$ be a physical relation among a
set of variable quantities $V_{i}$. The dimensional equations of
the different variables will be:

\begin{eqnarray*}
[V_{1}]&=&d_{1}^{\alpha_{11}}d_{2}^{\alpha_{12}}...d_{n}^{\alpha_{1n}}\\
\left [V_{2}\right ]&=&d_{1}^{\alpha_{21}}d_{2}^{\alpha_{22}}...d_{n}^{\alpha_{2n}}\\
...\\
\left [V_{n}\right
]&=&d_{1}^{\alpha_{m1}}d_{2}^{\alpha_{m2}}...d_{n}^{\alpha_{mn}}
\end{eqnarray*}

The matrix formed with the exponents of the dimensions is called
the Dimensional Matrix:

\begin{equation}\label{dimensional_matrix}
    \mathbf{D} = \left [ \begin{array}{cccc}
    \alpha_{11}&\alpha_{12}&\ldots&\alpha_{1n}\\
    \alpha_{21}&\alpha_{22}&\ldots&\alpha_{2n}\\
    \vdots&\vdots&\ldots&\vdots\\
    \alpha_{m1}&\alpha_{m2}&\ldots&\alpha_{mn} \end{array} \right ]
\end{equation}

From the dimensional matrix, we can now tackle the following
problem: how can we find dimensional (or dimensionless) products
of the variables at hand?. In other words, how do we find the
$\varepsilon_{1},\varepsilon_{2},...,\varepsilon_{n}$ exponents
that solve the following equation:

\begin{equation}\label{}
    [V_{1}^{\varepsilon_{1}}V_{2}^{\varepsilon_{2}}...V_{n}^{\varepsilon_{n}}]=
    d_{1}^{q_{1}}d_{2}^{q_{2}}...d_{p}^{q_{p}}
\end{equation}

where: $V_{1},...,V_{n}$ are the variables of the problem,
$d_{1},...,d_{p}$ are the fundamental dimensions of the problem,
$q_{1},...,q_{p}$ are the \emph{given} exponents (sought
combinations of fundamental dimensions; to be set equal to zero
for dimensionless products).

The problem is therefore reduced to solving for $\varepsilon_{j}$
the following system of linear equations:

\begin{equation}\label{exp_eqns}
    \sum_{j=1}^n \: D_{ij} \: \varepsilon_{j}\: =\: q_{i}
\end{equation}

It can be shown that (see \cite{SZIRTES1997}) the solution to the
above is:

\begin{equation}\label{solution_for_epsilon}
    \varepsilon_{j} \: = \:\sum_{k=1}^n \: E_{jk} \: \eta_{k}
\end{equation}

where:

\begin{displaymath}
\mathbf{E} = \left [ \begin{array}{cccc} \mathbf{1} & \vdots&\mathbf{0} \\
\ldots&&\ldots\\
-\mathbf{A^{-1}}\mathbf{B} &\vdots& \mathbf{A^{-1}}
\end{array} \right ]
\end{displaymath}

and $\mathbf{A}$ is a nonsingular square matrix of size the rank
of the dimensional matrix, and $\mathbf{B}$ is formed by the rest
of columns included in $\mathbf{D}$ and not included in
$\mathbf{A}$. The sizes of the different matrices are:

\begin{equation}\label{matrices}
\begin{array}{rlr}
\mathbf{E}:&(n)\times(n)&square \: matrix\nonumber\\
\mathbf{I}:&(n-p)\times(n-p)&identity \: matrix\nonumber\\
\mathbf{0}:&(n-p)\times(p) &null \: matrix\nonumber\\
\mathbf{A}:&(p)\times(p)&square \: matrix\nonumber\\
\mathbf{B}:&(p)\times(n-p)& \: matrix\\
\mathbf{A^{-1}B}:&(p)\times(n-p)&\: matrix\nonumber
\end{array}
\end{equation}

the column matrix $\eta_{k}$ is formed by $\varepsilon_{n-p}$
arbitrary exponents followed by $q_{p}$ sought dimensional
exponents. The first $\varepsilon_{n-p}$ are arbitrary due to the
fact that $R_{D}\equiv rank(\mathbf{D})=p$ and therefore out of
the n exponents we can only determine p, leaving n-p arbitrary.
Therefore, $\eta_{k}$ is:

\begin{displaymath}
\eta_{k}=\left [ \begin{array}{l}  \varepsilon_{1}\\
\varepsilon_{2}\\
\vdots\\  \varepsilon_{n-p}\ \\ q_{1} \\ q_{2}\\
\vdots \\ q_{p}\end{array} \right ]
\end{displaymath}

The problem is thus reduced to finding the matrix $\mathbf{P}$
result of the product:

\begin{equation}\label{exp_eqns}
    P_{ij}\: = \: \sum_{k=1}^n \: E_{ik} \: Z_{kj}\:
\end{equation}

where $E_{}ik$ is as defined earlier and $Z_{kj}$ is the result of
superimposing the $\eta_{k}$ columns for the different variables:

\begin{equation}\label{zeta_matrix}
    \mathbf{Z} = \left [ \begin{array}{cccc}
    \varepsilon_{11}&\varepsilon_{12}&\ldots&\varepsilon_{1p}\\
    \varepsilon_{21}&\varepsilon_{22}&\ldots&\varepsilon_{2p}\\
    \vdots&\vdots&\ldots&\vdots\\
    \varepsilon_{(n-p)1}&\varepsilon_{(n-p)2}&\ldots&\varepsilon_{(n-p)p}\\
    q_{1}&q_{1}&\ldots&q_{1}\\
    q_{2}&q_{2}&\ldots&q_{2}\\
    \vdots&\vdots&\ldots&\vdots\\
     q_{p}&q_{p}&\ldots&q_{p}
     \end{array} \right ]
\end{equation}

both $\mathbf{P}$ and $\mathbf{Z}$ are of size nxp.

In the particular case where we are seeking for
\emph{dimensionless} products, all the $q_{i}$ will be set to zero
(dimensional exponents) as a \emph{dimensionless} quantity is
considered to have dimension=1 (for a dissertation on the
convenience to call dimensionless these type of quantities, check
\cite{SZIRTES1997}).

\subsection{Complete set of dimensional products: Buckingham theorem}
The so called Buckingham's theorem reads as follows:

\emph{The number of independent dimensional products $n_{\pi}$
which can be composed for a given number of variables and
dimensions is:}
\begin{displaymath}
\begin{array}{rlll}
n_{\pi}\:&=&n\:-\:R_{D}&for \:dimensionless\:products\\
n_{\pi}\:&=&n\:-\:R_{D}\:+\:1&for \:dimensional\:products
\end{array}
\end{displaymath}

where $n$ is the number of variables and $R_{DM}$ is the rank of
the dimensional matrix.

In this case, the sizes of the matrices at \label{matrices} would
be:

\begin{equation}\label{matrices}
\begin{array}{rlr}
\mathbf{I}:&(n-R_{D})\times(n-R_{D})&identity \: matrix\nonumber\\
\mathbf{0}:&(n-R_{D})\times(R_{D})&null \: matrix\nonumber\\
\mathbf{A}:&(R_{D})\times(R_{D})&square \: matrix\nonumber\\
\mathbf{B}:&(R_{D})\times(n-R_{D})& \: matrix\\
\mathbf{E}:&(n)\times(n)&square \: matrix\nonumber\\
\mathbf{A^{-1}B}:&(R_{D})\times(n-R_{D})& \: matrix\nonumber
\end{array}
\end{equation}

An obvious advantage of finding the dimensionless products of the
problem under investigation is that it reduces the number of
parameters relevant to the system. In the case that the precise
mathematical formula governing the behavior of the system is
known, e.g., the Navier-Stokes equation for a fluid system, then
the dimensionless products give information as well on the
relative importance of each of the terms in the equation, helping
in the simplification of the equations for certain combinations of
the dimensionless products (for a beautiful example regarding the
Navier-Stokes equations, see \cite{ROCHE1980}).

\section{Application to Spectral Fluxes}
Using the dimensional analysis principles, we will extract the
dimensionless products for flux densities. These products will be
used later, in section 4, to achieve the unit conversion between
flux densities in different units.

\subsection{$F_{\lambda}$ versus $\lambda$}
To construct the dimensional matrix for $F_{\lambda}$ we will
consider a spectral energy distribution of the form:

\begin{equation}\label{f_l_relation}
F_{\lambda}(c,\lambda,h)=0
\end{equation}

The dimensional equations of the different members of the previous
relation are:

\begin{equation}\label{fl_vs_l_dim}
    [F_{\lambda}]=M L^{-1}T^{-3}, \;
    [c]=LT^{-1},\;
    [\lambda]=L,\;
    [h]=ML^2T^{-1}
\end{equation}

we now construct the following table:

\vspace{0.5cm}

\begin{equation}\label{f_l_vs_l}
\begin{tabular}{ l |  c c c c }
    & $F_{\lambda}$ & $c$ & $\lambda$ & $h$\\
  \hline
  M  & 1 & 0 & 0 & 1 \\
  L  & -1 & 1 & 1 & 2 \\
  T  & -3 & -1 & 0 & -1 \\
\end{tabular}
\end{equation}

according to previous sections, we will have the following set of
matrices:

\begin{eqnarray*}
    \mathbf{A} =\left[%
\begin{array}{rrr}
  0 & 0 & 1 \\
  1 & 1 & 2 \\
  -1 & 0 & -1 \\
\end{array}%
\right] \hspace{2cm}
 \mathbf{B} =\left[%
\begin{array}{r}
  1 \\
  -1 \\
  -3 \\
\end{array}%
\right]
\end{eqnarray*}

\begin{eqnarray*}
\mathbf{-A^{-1}B} =\left[%
\begin{array}{r}
  -2 \\
  5 \\
  -1 \\
\end{array}%
\right] \hspace{2cm} \mathbf{E} =\left[%
\begin{array}{rrrr}
  1 & 0 & 0 & 0 \\
  -2 & -1 & 0 & -1 \\
  5 & -1 & 1 & 1 \\
  -1 & 1 & 0 & 0 \\
\end{array}%
\right]
\end{eqnarray*}

\begin{eqnarray*}
\mathbf{Z} =\left[%
\begin{array}{rrrr}
  1 & 1 & 1 & 1 \\
  0 & 0 & 0 & 0 \\
  0 & 0 & 0 & 0 \\
  0 & 0 & 0 & 0 \\
\end{array}%
\right]\hspace{0.5cm} \mathbf{P} = \mathbf{E}*\mathbf{Z}=\left[%
\begin{array}{rrrr}
  1 & 1 & 1 & 1 \\
  -2 & -2 & -2 & -2 \\
  5 & 5 & 5 & 5 \\
  -1 & -1 & -1 & -1 \nonumber\\
\end{array}%
\right]
\end{eqnarray*}

where following the general practice (see \cite{SZIRTES1997}) we
have chosen the matrix A to be composed of the independent
variables in relation (\ref{f_l_relation}), and he matrix
$\mathbf{P}$ gives the exponents(columns) on the dimensions (rows)
that give rise to dimensionless products. According to the theory
before, the number of dimensionless products in this case would be
$n-R_{DM}$, i.e., 4-3=1, hence the fact that three of the columns
are linear combinations of the remaining one (identical, in this
case) in the P matrix only giving result to one dimensionless
product combination:

\begin{eqnarray*}
\begin{array}{lr}
\mathbf{P} =\left[%
\begin{array}{rrrr}
  1 & 1 & 1 & 1 \\
  -2 & -2 & -2 & -2 \\
  5 & 5 & 5 & 5 \\
  -1 & -1 & -1 & -1 \\
\end{array}%
\right]
&

\end{array}
\end{eqnarray*}

where rows identify the variables in (\ref{fl_vs_l_dim}) and
columns correspond to the exponents of those variables.

Therefore, the unique dimensional product thus for this case would
be:

\begin{equation}\label{dimless_fl_vs_l}
\pi = F_{\lambda}\frac{\lambda^{5}}{c^2 h}
\end{equation}

\subsection{$F_{\nu}$ versus $\lambda$}

Following exactly the same procedure as before, we would have:

\vspace{0.5cm}

\begin{equation}\label{f_n_vs_l}
\begin{tabular}{ l |  c c c c }
    & $F_{\nu}$ & $c$ & $\lambda$ & $h$\\
  \hline
  M  & 1 & 0 & 0 & 1 \\
  L  & 0 & 1 & 1 & 2 \\
  T  & -2 & -1 & 0 & -1 \\
\end{tabular}
\end{equation}

and therefore:

\begin{eqnarray*}
    \mathbf{A} =\left[%
\begin{array}{rrr}
  0 & 0 & 1 \\
  1 & 1 & 2 \\
  -1 & 0 & -1 \\
\end{array}%
\right] \hspace{2cm}
 \mathbf{B} =\left[%
\begin{array}{r}
  1 \\
  0 \\
  -2 \\
\end{array}%
\right]
\end{eqnarray*}

\begin{eqnarray*}
\mathbf{-A^{-1}B} =\left[%
\begin{array}{r}
  -1 \\
  3 \\
  -1 \\
\end{array}%
\right] \hspace{2cm} \mathbf{E} =\left[%
\begin{array}{rrrr}
  1 & 0 & 0 & 0 \\
  -1 & -1 & 0 & -1 \\
  3 & -1 & 1 & 1 \\
  -1 & 1 & 0 & 0 \\
\end{array}%
\right]
\end{eqnarray*}

\begin{eqnarray*}
\mathbf{Z} =\left[%
\begin{array}{rrrr}
  1 & 1 & 1 & 1 \\
  0 & 0 & 0 & 0 \\
  0 & 0 & 0 & 0 \\
  0 & 0 & 0 & 0 \\
\end{array}%
\right]\hspace{0.5cm} \mathbf{P} = \mathbf{E}*\mathbf{Z}=\left[%
\begin{array}{rrrr}
  1 & 1 & 1 & 1 \\
  -1 & -1 & -1 & -1 \\
  3 & 3 & 3 & 3 \\
  -1 & -1 & -1 & -1 \nonumber\\
\end{array}%
\right]
\end{eqnarray*}

And therefore the only dimensionless product would be:

\begin{equation}\label{dimless_fn_vs_l}
\pi' = F_{\nu}\frac{\lambda^{3}}{c h}
\end{equation}

Both dimensionless quantities in (\ref{dimless_fl_vs_l}) and
(\ref{dimless_fn_vs_l}) are descriptions of the same physical
problem. As they are lineal in the dependent variable
($F_{\lambda}$ and $F_{\nu}$ respectively) we are able to conclude
that they will be equivalent, i.e., that we can write in this
case:

$\pi=\pi'$

and therefore:

\begin{equation}\label{relation_fl_fn}
    F_{\lambda}\frac{\lambda^{5}}{c^2 h} = F_{\nu}\frac{\lambda^{3}}{c
    h} \:\:\Longrightarrow \:\: F_{\lambda}=\frac{c}{\lambda^2} F_{\nu}
\end{equation}

which is the physical result expected when transforming flux
densities.

The dimensionless product obtained when repeating the above
procedure for $F_{\nu}$ and $F_{\lambda}$ with $\nu$ as the
independent variable give:

\begin{eqnarray*}
\pi=F_{\nu}\frac{c^{2}}{\nu^{3} h} \:\:\:\:\mbox{and}\:\:
\:\:\:\pi=F_{\lambda}\frac{c^{3}}{\nu^{5} h}
\:\:\:\mbox{respectively}
\end{eqnarray*}

This working example shows how the dimensional analysis can be
used in the handling of spectrum fluxes and unit transformation.
Much more complex problems can be tackled using this approach. For
a nice compilation of literally hundreds of examples, consult the
book by \cite{SZIRTES1997}.

 In what follows, we give an algorithm to bring the aforementioned
 ideas to practice when designing a client tool to handle Virtual
 Observatory spectra.

\subsection{Comment on apparent velocity as X-axis spectral coordinate}
In the paper by \cite{GREISEN2004}, the apparent radial velocity
is considered as one of the possible spectral coordinates. A whole
set of possible transformations between the different spectral
coordinates ("x-axes") is given together with their derivatives.
The paper deals spectral coordinate transformations, rather than
with flux transformations (on the "y-axis"). In this work, we deal
with energy densities which depend, generally, on $\lambda$ or
$\nu$.

Conversions in the x-axis to velocity space do need a central
reference wavelength (frequency) which will have to accompany the
metadata for the data. Once this value is known, the
transformation of the x-axis will just consist of a translation
plus a dilation, and the Y-axis will be kept as is. In the case
that the data are coming with velocity in the x-axis, they will
have to contain the reference lambda giving rise to those velocity
values, and the process to convert the data to wavelength values
would be simply inverted.

\section{Algorithmic approach and use in VOSpec}
In the Virtual Observatory context, there is a need to have a
standard protocol to make spectra accessible from different
projects in a simple way. The idea is to create a protocol for
spectra in the same line as the already standard Simple Image
Access Protocol (SIAP), that would allow for the creation of
on-the-fly Spectral Energy Distributions (SEDs) from heterogeneous
data sets.

This protocol has been called Simple Spectrum Access Protocol
(SSAP) and basically implements a two-step process:

\begin{itemize}
    \item In the first step, a cone search is done on available
    services and the match results are sent, together with metadata, \
    in a VO standard VOTable.
    \item In the second step, the pointers to the real data files
    (spectra) are called and data are retrieved.
\end{itemize}

The main problem faced when trying to create an SED using data
from different projects/formats is to compare data different
units. We should be able to transform spectral coordinate units
and fluxes to a common unit system.

Our proposed solution is to specify in the metadata (first SSAP
step) for every spectrum, the dimensional equation for the
spectrum axes, and use dimensional analysis to extract the
conversion formulae needed to go from one to the other.

To prove that this on-the-fly conversion was possible, we
developed a tool called VOSpec, able to request different SSAP
servers and produce a common SED from different spectra in
different formats from different projects.

The application is already available to the general public at:
\textbf{http://esavo.esa.int/vospec/} and it has been used for the
AVO demo which took place on Jan 25 \& 26, 2005 at ESAC. See
\cite{AVODEMO2005} for reference.

\begin{figure*}
\begin{center}
\includegraphics[width=9cm, angle=0]{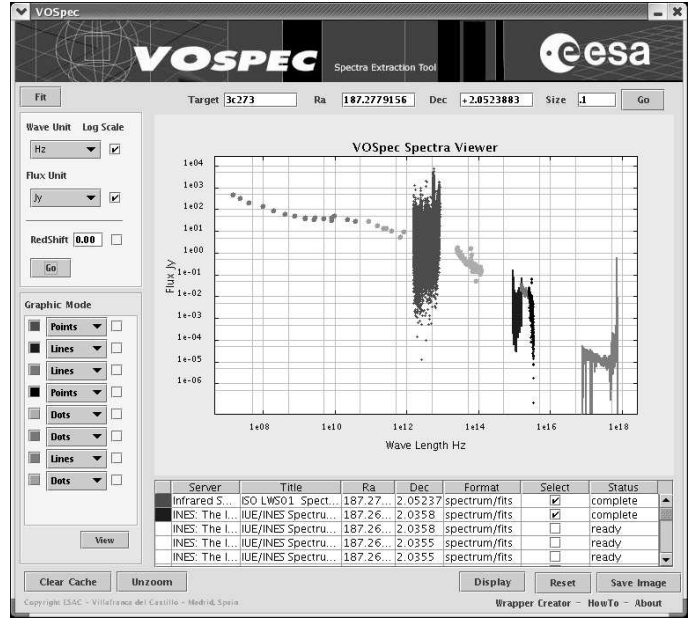}
\caption{SED generated for the quasar 3c273 using VOSpec, covering
from Radio to X-Ray spectral data. From left to right: Radio and
sub-milimiter points (local files loaded data taken from the
Geneve University database). Infrared spectrum data from the ISO
SSAP service (LWS and PHOT data). Optical and ultraviolet spectrum
data from FUSE (SSAP service) and HST(FOS) (SSAP service). On the
right, X-Ray data from XMM-Newton (processed local file)}
 \label{fig:VOSpec}
\end{center}
\end{figure*}


At the time of writing this paper, spectra from \textbf{ISO},
\textbf{IUE}, \textbf{HST (FOS)}, \textbf{SDSS},
\textbf{HyperLeda} and \textbf{FUSE} projects are already
providing SSAP access to their data. All the spectra from these
services can be superimposed in the same display, and the user can
generate on-the-fly SEDs as can be seen in Figure
~\ref{fig:VOSpec}. Spectra from projects that don't have SSAP
services, can be loaded locally using the SSAP Wrapper Creator
integrated in VOSpec.

\subsection{Algorithmic approach}
There are several ways to describe a spectrum flux and the
spectral coordinate inside a 1-D spectrum. A conversion table
could be used to make the transformations accordingly, but it is
not easy to use in an automatic algorithm, and this would limit
the number of possible transformations allowed by the system.

In this section, we will describe an algorithmic way to approach
the units problem from the dimensional point of view and show how
this is used in the VOSpec application.\\

A unit can be described in the following way:

\begin{equation}
[UNIT] = Scaling * M^{a}L^{b}T^{c}
\end{equation}
where the scaling factor is defined with respect to a certain
common system of units and the exponents a,b,c define the unit
dimensionally. We will choose the SI as our base dimensional
system of reference.

In order to understand how the algorithm works, suppose we are
dealing with a spectrum in Jansky (y-axis) and Hertz (x-axis). The
units dimensions (and scale factors) turn out to be:

\begin{eqnarray*}
    \nu (Hz) &=& T^{-1}\\
    F_{\nu }(Jy) &=& 10^{-26} MT^{-2}
\end{eqnarray*}

where $1$ and $10^{-26}$ are the reference scaling factors to the
SI units system for $Hz$ and $Jy$ respectively.

Suppose now, we want to convert one spectrum point defined by the
pair $\left[\nu (Hz),F_{\nu }(Jy)\right]$ to other ones, e.g.
$\left[\lambda (\mu m), F_{\lambda }(\frac{W}{cm^{2} \mu
m})\right]$, i.e.:
\begin{eqnarray*}
    \lambda (\mu m) &=& 10^{-6} L \\
    F_{\lambda }(\frac{W}{cm^{2} \mu m}) &=& 10^{10} ML^{-1}T^{-3}
\end{eqnarray*}

First we need to generate the matrix $E$ for the original system,
as we saw in section 3:

\begin{equation}\label{f_l_vs_l}
\begin{tabular}{ l |  c c c c }
    & $F_{\nu }$ & $c$ & $\nu $ & $h$\\
  \hline
  M  & 1 & 0 & 0 & 1 \\
  L  & 0 & 1 & 0 & 2 \\
  T  & -2 & -1 & -1 & -1 \\
\end{tabular}
\end{equation}
so, in this case:
\begin{eqnarray*}
\mathbf{A} =
\left[%
\begin{array}{rrr}
0&0&1\\
1&0&2\\
-1&-1&-1\\
\end{array}%
\right] \hspace{0.5cm} \mathbf{A^{-1}} =
\left[%
\begin{array}{rrr}
-2&1&0\\
1&-1&-1\\
1&0&0\\
\end{array}%
\right]
\end{eqnarray*}

and constructing $\mathbf{B}$ from the Flux density units:

\begin{eqnarray*}
\mathbf{B} =
\left[%
\begin{array}{r}
1\\
0\\
-2\\
\end{array}%
\right] \hspace{0.5cm} \mathbf{-A^{-1} B} =
\left[%
\begin{array}{r}
2\\
-3\\
-1\\
\end{array}%
\right]
\end{eqnarray*}

so finally, constructing the matrix $\mathbf{E}$ using the rules
described in section 3:

\begin{eqnarray*}
\mathbf{E} =
\left[%
\begin{array}{rrrr}
1&0&0&0\\
2&-2&1&0\\
-3&1&-1&-1\\
-1&1&0&0\\
\end{array}%
\right]
\end{eqnarray*}

Now, to go from one system to the other, we have to generate two
different Z matrices, one for the spectrum coordinate
transformation and the second one for the flux transformation. As
we are looking for certain target dimensions, the Z matrix will
have to include the sought dimensional exponents, and therefore,
as the final spectrum coordinates are $\mu m$ ($L$) we will have:

\begin{eqnarray*}
\begin{array}{lr}
\mathbf{Z_{1}} =
\left[%
\begin{array}{r}
  0 \\
  0 \\
  1 \\
  0 \\
\end{array}%
\right]
&
\begin{array}{rrrr}
  F_{\nu }\\
  M \\
  L\\
  T\\
\end{array}%
\end{array}
\end{eqnarray*}

where the first zero is imposing no dependence in the flux, and
the rest are the exponents in $M$, $L$, $T$ respectively.

For the final system of units, the $F_{\lambda }$ is
$\frac{W}{cm^{2} \mu m}$ ($ML^{-1}T^{-3}$):

\begin{eqnarray*}
\begin{array}{lr}
\mathbf{Z_{2}} =
\left[%
\begin{array}{r}
  1 \\
  1 \\
  -1 \\
  -3 \\
\end{array}%
\right]
&
\begin{array}{rrrr}
  F_{\nu }\\
  M \\
  L\\
  T\\
\end{array}%
\end{array}
\end{eqnarray*}

where, as it was defined in previous section, the first element is
the dependence in the flux and the rest of the elements in the $Z$
are the dimensional equation exponents for $M$,$L$ and $T$
respectively.

If we multiply then the original E matrix with these Z vectors we
obtain for the spectral coordinate:

\begin{eqnarray*}
\begin{array}{lr}
\mathbf{P_1} = \mathbf{E  * Z_1} =
\left[%
\begin{array}{r}
  0 \\
  1 \\
  -1 \\
  0 \\
\end{array}%
\right]
&
\begin{array}{rrrr}
  F_{\nu }\\
  c \\
  \nu\\
  h\\
\end{array}%
\end{array}
\end{eqnarray*}

and for the flux:

\begin{eqnarray*}
\begin{array}{lr}
\mathbf{P_2} = \mathbf{E  * Z_2} =
\left[%
\begin{array}{r}
  1 \\
  -1 \\
  2 \\
  0 \\
\end{array}%
\right]
&
\begin{array}{rrrr}
  F_{\nu }\\
  c \\
  \nu\\
  h\\
\end{array}%
\end{array}
\end{eqnarray*}

That means, respectively:

\begin{eqnarray*}
\left[ \lambda \right] &=& \frac{\left[c\right]}{\left[ \nu \right] } \\
\left[ \frac{W}{cm^{2} \mu m} \right] &=& \left[ Jy\right] .
\frac{\left[\nu \right] ^2}{\left[c\right]}
\end{eqnarray*}

To finalize the transformation, we must include the scaling
factors of the different units. To achieve this goal, we note that
every time a magnitude is used, the scaling must appear, i.e.,

\begin{eqnarray*}
S_{(\mu m)} \lambda (\mu m)&=& \frac{c}{S_{(Hz)} \nu (Hz)} \\
S_{(\frac{W}{cm^{2} \mu m})} F_{\lambda }(\frac{W}{cm^{2} \mu m})
&=& S_{(Jy)} F_{\nu }(Jy) . \frac{(S_{(Hz)} \nu (Hz))^2}{c}
\end{eqnarray*}

Where the S elements correspond to the scalings with respect to a
common system of reference units (in this case SI).\\

Finally we obtain:

\begin{eqnarray*}
\lambda (\mu m)  &=& \frac{1}{S_{(\mu m)}.S_{(Hz)}}
.\frac{c(m/s)}{\nu (Hz)} \\
&=& 10^6 . \frac{c(m/s)}{\nu (Hz)} \\
F_{\lambda }(\frac{W}{cm^{2} \mu m}) &=& \frac{S_{(Jy)} .
S^2_{(Hz)}}{S_{(\frac{W}{cm^{2} \mu m})}} F_{\nu }(Jy) \frac{\nu
^2
(Hz)}{c(m/s)} \\
&=& \frac{10^{-26}}{10^{10}} F_{\nu }(Jy) \frac{\nu ^2
(Hz)}{c(m/s)} \\
&=& 10^{-36} F_{\nu }(Jy) \frac{\nu ^2 (Hz)}{c(m/s)}
\end{eqnarray*}

These final formulae tell us how to express the values of a
$\left[x,y\right]$ point in the final units as a function of the
point in the original ones.

The algorithm can be summarized as follows:

\begin{enumerate}
    \item Construct the matrix A, using the spectral coordinate, c and h
    dimensional equations.
    \item Construct the vector B using the flux density dimensional equation.
    \item Invert the matrix A, and construct the matrix E as it
    was described in point 1.
    \item Construct two Z vectors using spectral coordinate and flux
    density dimensional equations of the final units.
    \item Multiply the matrix E with the two Z vectors to obtain
    the conversion factors.
    \item Finally, use the scaling factors to finish the
    conversion.
\end{enumerate}

\section{Conclusions} We have shown how to make use of
Dimensional Analysis techniques to handle unit conversion in an
automated way for the case of spectral flux densities.

We have proposed the IVOA to include the SCALEQ and DIMEQ
parameters as part of the Simple Spectrum Access Protocol so that
clients can use Dimensional Analysis algorithms to handle units
automatically.

The approach shown is not only relevant to the Spectral access
within the VO, and can be extended -as mentioned in the
introduction- to any physical problem by including the relevant
dimensions.

In this sense, we imagine any unit as composed of three main
attributes:

\begin{enumerate}
\item A name (and possibly a symbol)
\item A SCALEQ (giving the
Scaling of the unit with respect to the International System of
Units)
\item A DIMEQ (giving the dimensions of the unit)
\end{enumerate}

An example of a serialization of this units model could be the
FITS representation, in which a unit like the "Jansky" could be
represented in the following way:

\begin {eqnarray*}
\begin{array}{lll}
CUNIT='Jansky' \\
CSCALEQ='1E-26' \\
CDIMEQ='MT**-2'
\end{array}
\end{eqnarray*}

Certainly, to cover all possible physical units, more dimensions
than M-L-T would be needed, but the set will always be far smaller
than the units they can represent.

\vspace{5cm}

\begin{acknowledgements} We acknowledge Jose Tomas
Diez-Roche, Matteo Guainazzi and Andy Pollock for their useful
comments and discussions.

\end{acknowledgements}

\end{document}